\documentclass[runningheads]{llncs}
\usepackage{graphicx}
\usepackage{amsmath}
\usepackage{url}
\usepackage{subfigure}
\usepackage{booktabs}
\usepackage{caption}
\usepackage{hyperref}
\usepackage[skip=2pt]{caption}
\newcommand{\distance}{3pt}
\begin{document}
\title{An Empirical Study on JIT Defect Prediction Based on BERT-style Model}
%
%
\author{Yuxiang Guo\inst{1}\ \and
Xiaopeng Gao\inst{1} \and
Bo Jiang\inst{1,*}}
\authorrunning{Yuxiang Guo et al.}
%

\institute{State Key Laboratory of Software Development Environment, School of Computer Science,
Beihang University, Beijing, China \\
\email{\{irisg,gxp,bojiang\}@buaa.edu.cn}\\ 
 \inst{*} Corresponding author\\}

\index{SurnameAuthor1, FirstnameAuthor1}

%
\maketitle              

\begin{abstract}
Previous works on Just-In-Time (JIT) defect prediction tasks have primarily applied pre-trained models directly, neglecting the configurations of their fine-tuning process. In this study, we perform a systematic empirical study to understand the impact of the settings of the fine-tuning process on BERT-style pre-trained model for JIT defect prediction. Specifically, we explore the impact of different parameter freezing settings, parameter initialization settings, and optimizer strategies on the performance of BERT-style models for JIT defect prediction. Our findings reveal the crucial role of the first encoder layer in the BERT-style model and the project sensitivity to parameter initialization settings. Another notable finding is that the addition of a weight decay strategy in the Adam optimizer can slightly improve model performance. Additionally, we compare performance using different feature extractors (FCN, CNN, LSTM, transformer) and find that a simple network can achieve great performance. These results offer new insights for fine-tuning pre-trained models for JIT defect prediction. We combine these findings to find a cost-effective fine-tuning method based on LoRA, which achieve a comparable performance with only one-third memory consumption than original fine-tuning process.
\end{abstract}
\section{Introduction}
Defect prediction aids developers in identifying software defects before deploying potentially faulty code. Just-In-Time (JIT) defect prediction, as discussed by researchers like ~\cite{2008Classifying}, traditionally relies on manually extracted features and statistical machine learning methods, such as logistic regression and SVM ~\cite{9463103,9026802}.

With the advancement of deep learning technology, some researchers construct JIT defect prediction models with end-to-end techniques using the source code as input. For example, Hong et al. ~\cite{8816772} propose the DeepJIT technique to extract features directly from commit code and commit message submitted by the developer. CC2Vec improves DeepJIT with a pre-trained distributed change-code vector, resulting in performance improvements ~\cite{9284081}. Zeng et al. ~\cite{issta21} propose a JIT defect prediction method named LApredict, which only uses the number of added lines in one commit data as input with logistic regression as the classifier.

The emergence of pre-trained models introduces a new two-step paradigm of pre-training and fine-tuning to address typical tasks. Some transformer encoder-based pre-trained models like RoBERTa ~\cite{zhuang-etal-2021-robustly}, CodeBERT ~\cite{feng-etal-2020-codebert}, decoder-based pre-trained models like GPT2 ~\cite{Radford2019LanguageMA}, CodeGPT, encoder-decoder based model like BART ~\cite{lewis-etal-2020-bart} and PLBART ~\cite{plbart} have been proposed. Researchers applying pre-trained models to programming language tasks, such as text classification, code search, and vulnerability detection, have shown significant improvements ~\cite{Linevul,Commitdetection,icse23,zhang2023gamma}. 

BERT-style models are the variants of BERT model proposed in Devlin et al.'s work ~\cite{devlin-etal-2019-bert}. Some researchers explore the original BERT model through some other optimized methods. RoBERTa improved the pretraining procedure by exploring design decisions when pretraining BERT ~\cite{devlin-etal-2019-bert}, CodeBERT ~\cite{feng-etal-2020-codebert} is one variant of RoBERTa using natural language and programming language which can be adapted into code-related task. These BERT-style models provide new cornerstones to solve a certain problem. 

However, most previous researchers only focus on the application of pre-trained models in downstream tasks, but how different factors affect task performance is still unknown. Although Zhang et al. ~\cite{zhang2021revisitingbert} explored the fine-tuning of BERT contextual representations, but they only focused on common natural language understanding tasks in few-shot scenario. For better understanding pre-trained model applied in JIT defect prediction task, it is crucial to systematically explore how different factors affect the performance of the pre-trained model for JIT defect prediction task. 

In this study, to comprehensively evaluate the factors affecting the fine-tuning of the two BERT-style models for JIT defect prediction, we conduct an empirical study and select two effective BERT-style models CodeBERT and RoBERTa which are used in previous work for defect prediction ~\cite{Linevul,AssessingGeneralizabilityCodeBERT,guo2023study} and have obtained great performance in \textit{qt} and \textit{openstack} project. We utilized commit code and commit message as model inputs and evaluated the performance of these two models on a scale of nearly 100,000 extend defect prediction dataset ~\cite{issta21} to compare their performance under different fine-tune settings. Lastly, we combine our findings to propose a more efficient fine-tune method based on LoRA ~\cite{lora2021} to optimize CodeBERT for defect prediction with lower memory consumption.

The contribution of this work is fourfold. Firstly, our research delves deeply into the comparative analysis of different tuning factors in two BERT-style JIT defect prediction models. Secondly, we conduct an in-depth analysis of performance differences under various parameter freezing and initialization strategies across six projects. Our findings highlight the crucial role of the first encoder layer for model performance and demonstrate that appropriately adjusting weight initializing strategies can lead to performance improvements. Thirdly, we select different feature extractors in RoBERTaJIT and CodeBERTJIT to explore their impact on model performance, we find that simple network like FCN can achieve great performance than other complex structure like transformer. Finally, we propose a more efficient fine-tuning way based on our findings with the LoRA method. Our implementation of experiments is accessible at \href{https://github.com/AresXD/JIT-defect-prediciton-study-on-bert-style-model}{BERT style model on JTT}

The organizations of this work is as follows. In section ~\ref{sec:background}, we present the background and related work on JIT defect prediction and BERT-style models. In section ~\ref{sec:experiment}, we present the design and the results of our empirical study. Then, in section ~\ref{sec:threats}, we talk about threats and limitations to be verified. In section ~\ref{sec:conclusions}, we conclude our work. 
\section{Background and Related Work}
\label{sec:background}
In this section, we briefly revisit some related work on JIT defect prediction and fine-tune exploration based on pre-trained models.

\subsection{JIT defect prediction}
Finding defects in time can reduce the cost of defects and improve the quality of software. Researchers have proposed a large number of defect prediction techniques, which mainly include module-level, file-level, and change-level defect prediction based on different prediction granularity. In 2008, Kim et al. \cite{2008Classifying} proposed a code classification method using machine learning, i.e., code is classified into bug and bug-free categories. In 2013, Kamei et al. \cite{2013kamei} named the technique predicting defects based on software change characteristics as JIT defect prediction.

The features used in JIT defect prediction mainly include traditional features extracted manually and features extracted by deep learning methods. Mockus et al. proposed selecting manually extracted features for the first time and dividing them into different categories of features ~\cite{2000Predicting}, including changes in lines of code, number of files modified, and developer's experience. Some other works have extended the traditional artificial features by adding features such as code complexity, word frequency of logs or code, and the difference in the node number of abstract syntax tree before and after modification of relevant code files to represent changes ~\cite{2008Classifying,7202954,2016Studying,8453123,6772130,2003Ordering,1542070}.

Depending on the development of deep learning technology in recent years \cite{cnn,lstm}, the proposal of various network structures provides new ideas for feature extraction. Hong et al. \cite{8816772} introduced a novel deep Just-In-Time (JIT) defect prediction model, utilizing commit code and commit message as the model input based on CNN network.CC2Vec \cite{9284081} can learn representation of code changes by pre-training the Commit code into a distributed vector in advance. Hong et al. achieved performance improvements in the JIT defect prediction task by concatenating the CC2Vec vector with other input data from DeepJIT.

Zeng et al. \cite{issta21} proposed six new datasets from qt, openstack, jdt, platform, their work proposed a JIT defect prediction method named LApredict, which only uses the number of added line in one commit data as input with logistic regression as the classifier. Zhou et al. \cite{AssessingGeneralizabilityCodeBERT} used both commit code and commit message to construct a CodeBERT-based model, and evaluated the performance of the model on qt and openstack projects. Inspired by Zhou et al. \cite{AssessingGeneralizabilityCodeBERT}, Guo et al. \cite{guo2023study} did a more complete exploration of the JIT defect prediction task with different pre-trained models as the backbone.

\subsection{Tuning pre-trained BERT-style models}
Pre-trained models rely on the proposed transformer structure \cite{transformers}. One significant category of pre-trained models is transformer-encoder-based models, among them the BERT-style models are the most popular, such as BERT \cite{devlin-etal-2019-bert}, RoBERTa \cite{zhuang-etal-2021-robustly} and CodeBERT \cite{feng-etal-2020-codebert}. BERT-style models have the same transformer encoder stacked structure that was first proposed in Devlin et al.'s work \cite{devlin-etal-2019-bert}. 

Pre-trained models can be adapted to new tasks through fine-tuning. These models have gained prominence in natural language processing \cite{bertstyledifference,BERTstyleaugmenting} tasks. They have also demonstrated strong performance in various tasks within the field of software engineering and security \cite{Linevul,zhang2023gamma}. Initially, researchers often opt for full fine-tuning on a new downstream task. The instability of the BERT fine-tuning process has been acknowledged since its introduction \cite{devlin-etal-2019-bert}, leading to the proposal of various methods to address this issue. Phang et al. \cite{Stability} demonstrated that fine-tuning the pre-trained model on a large intermediate task stabilizes later fine-tuning on small datasets. Lee et al. \cite{rifleLi} introduced a new regularization method to constrain the fine-tuned model to stay close to the pre-trained weights, showcasing its stabilizing effect on fine-tuning. Mosbach et al. \cite{optimizer} showed that BERTADAM leads to instability during fine-tuning.

In contrast to exploring tuning stability during the full fine-tuning scenario, Aghajanyan et al. \cite{aghajanyan-etal-2021-intrinsic} found that larger networks or networks pre-trained on more data require smaller modifications in terms of the rank of the range to learn a new task. This observation explains the success of parameter-efficient fine-tuning (PEFT) methods and has motivated the development of low-rank fine-tuning methods such as LoRA \cite{lora2021} and Compacter \cite{Compactor}. Unlike the full fine-tuning paradigm, parameter-efficient fine-tuning methods are based on the assumption that achieving the same performance as the entire network is possible in the early stages of complex network training. 
\section{Empirical Study}
\label{sec:experiment}
In this section, we present details of our empirical study to investigate the factors impacting the fine-tuning process of the BERT-style model. 

\subsection{Research Questions}
In this section, we outline the research questions to explore in this work.

{\bfseries RQ1}: How do parameter freezing and initialization impact BERT-style model performance?

{\bfseries RQ2}: To what extent are the BERT-style models sensitive to different feature extractors? 

{\bfseries RQ3}: What is the impact of the optimizer strategies on BERT-style models?

{\bfseries RQ4}: Can we find a cost-effective fine-tune configuration for a BERT-style model based on the findings of previous RQs?
\subsection{Datasets}
To facilitate the evaluation, we borrow the dataset from Zeng et al.'s work \cite{issta21}, which is collected from 6 popular open-source projects (i.e., qt, openstack, jdt, platform, gerrit and go) on GitHub with the help of manual analysis and labeling by SZZ algorithm \cite{szz1,szz2}. We selected the data in the last three years except for the jdt project, in which we used a longer data period because it has fewer data items than other projects. The dataset used for evaluation is shown in Table \ref{tab:dataset}, which consists of 99,412 labeled commits. The Commit column is the total amount of commit patches of the corresponding project. And Defect column is the amount of defective commit patches in each project. \%Defect column is the defective rate of commit patches in each project. Among them, 26,693 of them are security-related patches and the left are non-security patches. We split the data of each project with the ratio of 80:20, 80\% of the data is used as the training set, and the remaining 20\% of the data is used for testing.
\begin{table}\small
\setlength{\tabcolsep}{4pt}
\centering
  \caption{Statistics of the Dataset}
  \label{tab:dataset}
  \begin{tabular}{llll}
 \hline
    {\bf{Project}} & \bf{Commit} & \bf{Defect} & \bf{\%Defect}\\ 
 \hline 
   qt & 23912 &3582&15.0\\
   openstack &22757 &5927&26.0\\
   jdt &7773 &3202&41.2\\ 
platform &11034 &4056&36.8\\
gerrit &14927 &1781&11.9\\ 
go &19009 &8145&42.8\\ \hline
all&99412 &26693&26.9\\ \hline
\end{tabular}
\end{table}
\subsection{Experiment Setup}
We use a desktop PC as our experiment environment. The PC is running Ubuntu 18.04 and is equipped with RTX3090 with 24GB of memory. To reduce the effect of randomness, all the experiments are repeated five times to obtain the mean value. In this work, all RQs is trained in 3 epochs and learning rate is set into 1e-5, batch size is set into 16. We follow the model structure and parameter setting of Zhou et al.'s ~\cite{AssessingGeneralizabilityCodeBERT} work, each commit code contains four commit files. In RQ1 and RQ3, we keep the CNN feature extractor the same as Zhou et al.'s ~\cite{AssessingGeneralizabilityCodeBERT} work. We only use LoRA in RQ4, LoRA rank is set into 8, and the first encoder layer of CodeBERT is not frozen.

\subsection{JIT defect prediction models for Comparison}
\textbf{DeepJIT:} DeepJIT \cite{8816772} is an end-to-end JIT defect prediction model utilizing commit code and commit message as the model input and utilizes CNN to extract meaningful features to achieve classification tasks. Evaluated in qt and openstack projects, DeepJIT can obtain a better performance than other traditional machine learning JIT defect prediction methods.

\textbf{CC2Vec:} To find a suitable vector representation with generalization, Hong et al. \cite{9284081} use the LSTM network to pre-train distributed vector CC2Vec which can learn a representation of code changes learning to commit message information in advance. The combination of CC2Vec and other input data from DeepJIT brings performance improvements in the JIT defect prediction task in both qt and openstack projects.

\textbf{LApredict:} Zeng et al. \cite{issta21} proposed a JIT defect prediction method named LApredict, which only uses the number of added code lines in one commit data as input with logistic regression as classifier. 

\textbf{BERT-style-based JIT defect prediction models:} Zhou et al. \cite{AssessingGeneralizabilityCodeBERT} use both commit code and commit message to construct a CodeBERT-based model and evaluate the performance of the model on qt and openstack projects. Inspired by Zhou et al., Guo et al. \cite{guo2023study} did a more complete exploration of the JIT defect prediction task with different pre-trained models as the backbone and explore both RoBERTa and CodeBERT as the model backbone. We select RoBERTa and CodeBERT as two backbones to construct RoBERTaJIT and CodeBERTJIT, which are two BERT-style based JIT defect prediction models. We utilize both commit code and commit message as model inputs.

\begin{figure}[h]
    \centering
    \includegraphics[width=0.80\textwidth ]{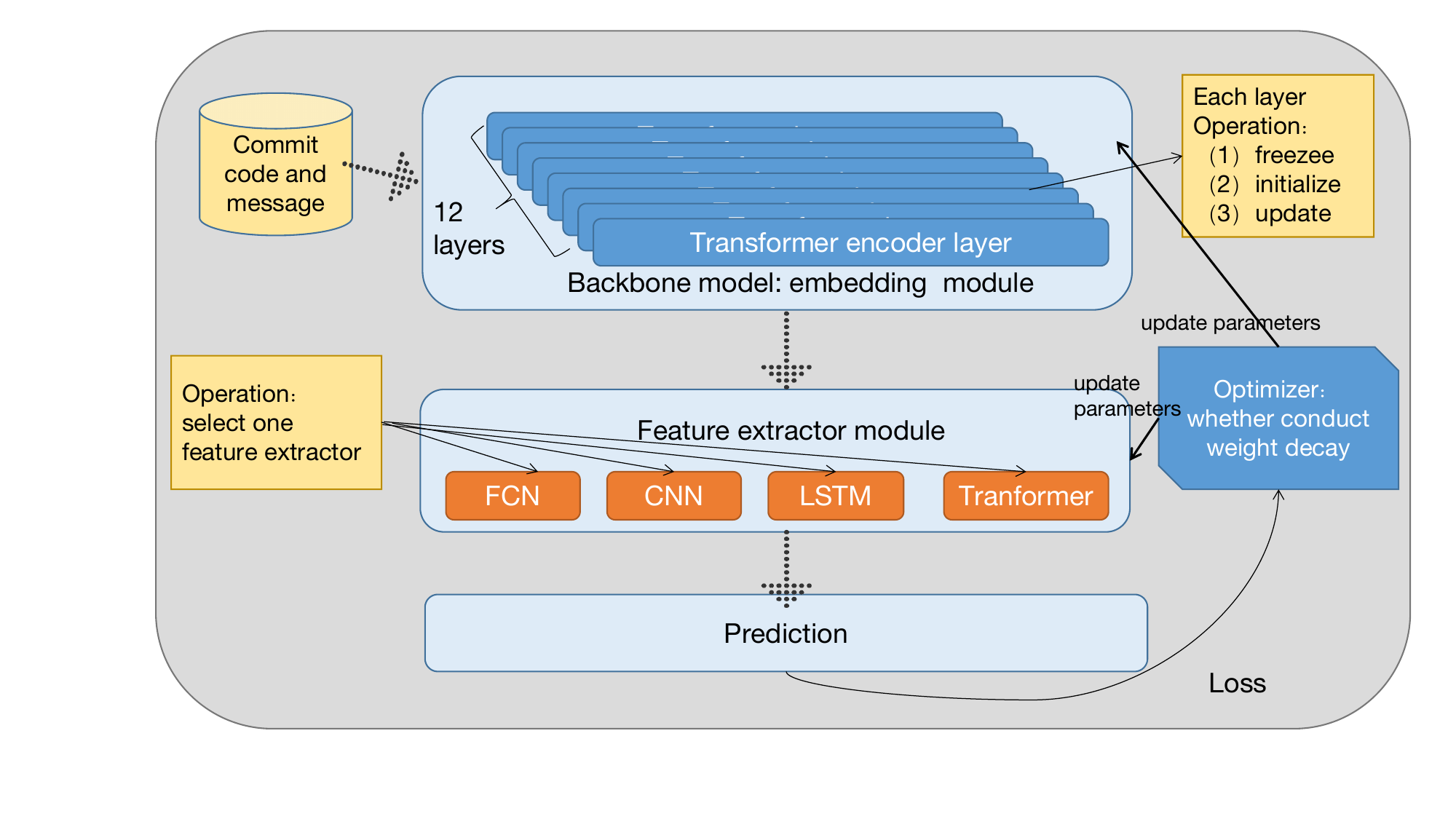}
    \caption{Various methods adjustable in the fine-tuning process of the BERT-style JIT defect prediction models.}
    \label{fig:framework}
\end{figure}

The settings to fine-tune the RoBERTaJIT and CodeBERTJIT models can be categorized into three parts: the embedding module setting, the extractor module setting, and the optimizer setting. Figure \ref{fig:framework} illustrates the methods that can be adjusted during the fine-tuning process of these BERT-style JIT defect prediction models.

In the embedding module, which essentially consists of stacked transformer encoder layers, three different operations can be performed. These operations include initializing the weight parameters with a normal distribution before fine-tuning, freezing the weight parameters during fine-tuning, and updating the weight parameters. In the feature extractor module, four common feature extraction networks are selected: FCN, CNN (which is used in the original RoBERTaJIT and CodeBERTJIT models), LSTM, and Transformer. In the optimizer module, two choices are provided for the Adam optimizer: common Adam and Adam with a weight decay strategy. This framework allows for a comprehensive exploration of how different factors influence the BERT-style JIT defect prediction model.
\subsection{Evaluation Metrics}
Four different results are produced when a binary model is used for prediction. We utilize the metrics including AUC score and F1-score. These indicators are commonly used in the literature to evaluate performance in the defect prediction task ~\cite{Linevul,guo2023study,AssessingGeneralizabilityCodeBERT,issta21,8816772,9284081}. Both AUC scores and F1-scores range from 0 to 1, all these two metrics have the best value of 1. And For each research question results, We conduct T tests on all predictions distribution considering the p-value less than 0.05.

\subsection{Experiment and Results Analysis}
Our analysis is based on the results that the predicted distribution of T-test results between different settings is statistically different considering p-value less than 0.05. All p-value results are less than 0.05 except for p-values greater than 0.05 between RNN and other feature extractors in RoBERTaJIT model and between reinitializing first 3 encoder layers and first 4 encoder layers in CodeBERTJIT model.

{\bfseries Results and analysis for RQ1.} Previous defect prediction work based on pre-trained models tends to conduct full fine-tuning operations. How different parameter settings affect the performance of defect prediction is still unknown. We studied two different parameter setting paradigms for fine-tuning: freezing part of model parameter and initializing parameters before fine-tuning a downstream task. We also repeat the experiment 5 times to handle the impact of randomness.
\begin{figure}[h]
    \centering
    \subfigure[RoBERTaJIT]{
     \includegraphics[width=0.46\textwidth]{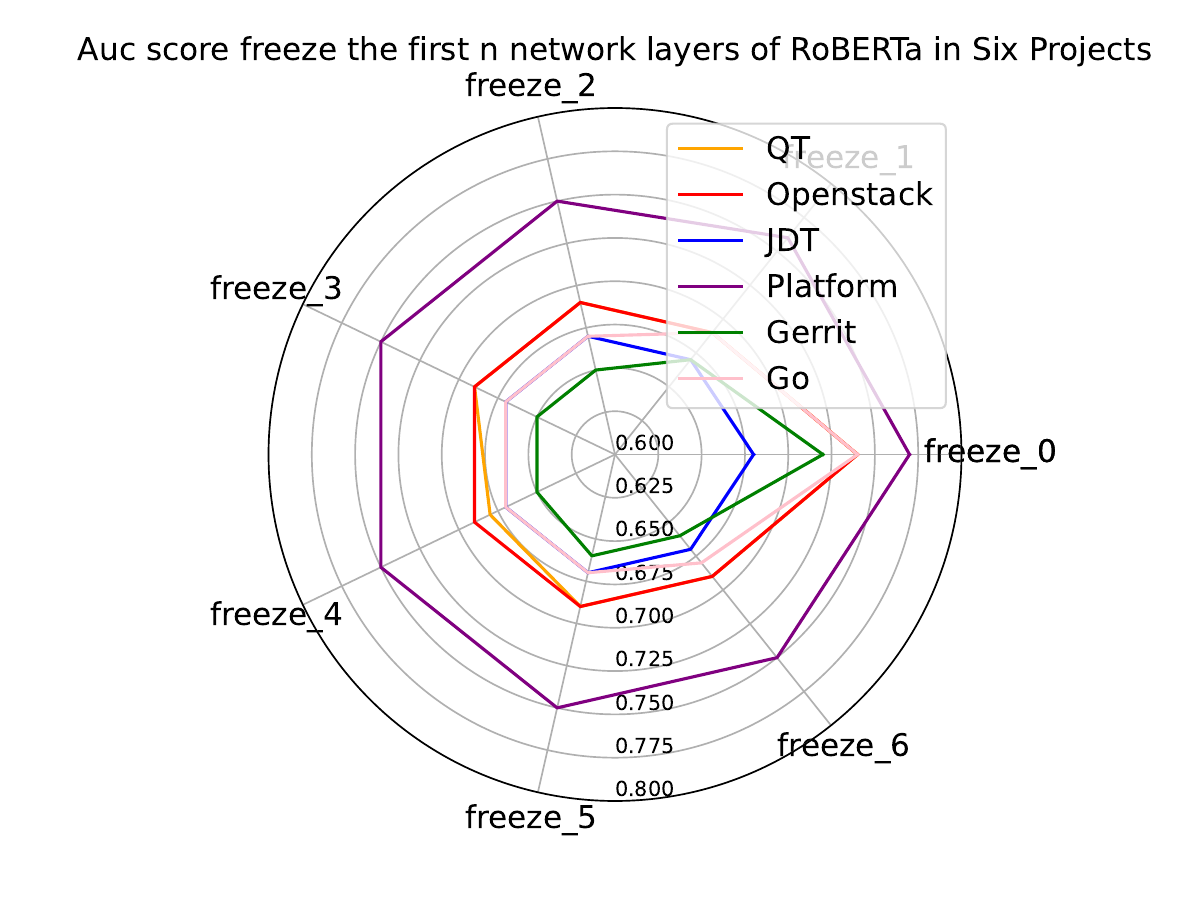}
    }
    \subfigure[CodeBERTJIT]{
       \includegraphics[width=0.46\textwidth ]{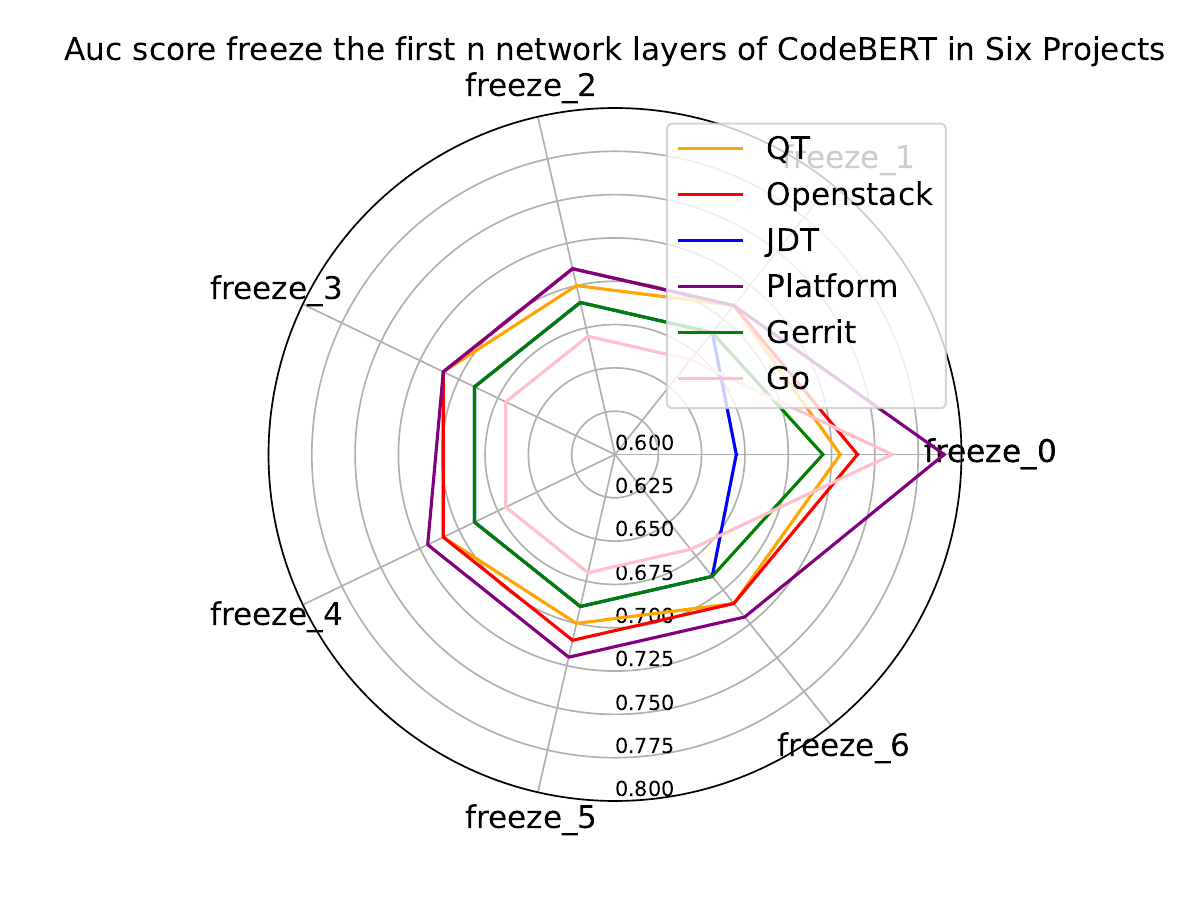}
    }
    \caption{AUC change with different freezing settings in RoBERTaJIT and CodeBERTJIT}
    \label{fig:freeze}
\end{figure}
\begin{table}\small
\setlength{\tabcolsep}{4pt}
  \caption{F1 score of freezing first n layers in RoBERTaJIT model}
  \centering
  \label{tab:roberta_freeze_F1}
  \begin{tabular}{cccccccc}
    \hline
    \bf{Project} & \bf{$f_0$} & \bf{$f_1$} & \bf{$f_2$}& \bf{$f_3$}&\bf{$f_4$}&\bf{$f_5$}&\bf{$f_6$}\\
    \hline
qt&0.37 & 0.33 & 0.33 & 0.33 & 0.33 & 0.33 & 0.34\\
openstack&0.42 & 0.39 & 0.39 & 0.39 & 0.39 & 0.39 & 0.41\\
jdt&0.62 & 0.56 & 0.57 & 0.57 & 0.57 & 0.57 & 0.62\\
platform&0.60 & 0.51 & 0.54 & 0.53 & 0.53 & 0.54 & 0.5\\
gerrit&0.29 & 0.2 & 0.2 & 0.2 & 0.2 & 0.2 & 0.2\\
go&0.54 & 0.54 & 0.55 & 0.52 & 0.53 & 0.57 & 0.52\\ \hline
mean & \textbf{0.47} & 0.42$\downarrow$ & 0.43$\downarrow$ & 0.42$\downarrow$ & 0.42$\downarrow$ & 0.43$\downarrow$ & 0.43$\downarrow$\\
\hline
\end{tabular}
\end{table}
We freeze the first n encoder layers in the backbone model (i.e., RoBERTa and CodeBERT), then we train and evaluate the RoBERTaJIT and CodeBERTJIT models on six projects. The experimental results for freezing different encoder layers in RoBERTaJIT and CodeBERTJIT are depicted in Figure \ref{fig:freeze}, Table \ref{tab:roberta_freeze_F1}, and Table \ref{tab:codebert_freeze_F1}. RoBERTa and CodeBERT model combines 12 transformer encoder layers. In the freeze setting, we freeze the first n layers of RoBERTa and CodeBERT where n varies from 0 to 6, freeze\_0 means we do not freeze any encoder layer. For brevity, we use $f_i$ as an abbreviation for freeze\_i in Table \ref{tab:roberta_freeze_F1} and Table \ref{tab:codebert_freeze_F1}.

As shown in Figure \ref{fig:freeze}, Table \ref{tab:roberta_freeze_F1}, and Table \ref{tab:codebert_freeze_F1}, after freezing the parameters of the first encoder layer, both RoBERTaJIT and CodeBERTJIT models experience a significant decline in AUC score and F1 score across all six projects. Interestingly, when freezing the parameters of the first n layers with n greater than 1, the AUC score performance does not exhibit a significant difference compared to freezing only the first layer. This phenomenon can be attributed to the accumulation of errors in the forward process during model training. The initial layers of a model are inclined to learn general knowledge relevant to the downstream task. Therefore, when fine-tuning a model based on a pre-trained model, updating parameters in the first layers enables the model to grasp information more pertinent to a certain task. 
\begin{table}\small
\setlength{\tabcolsep}{4pt}
  \caption{F1 score of freezing first n layers in CodeBERTJIT model}
  \centering
  \label{tab:codebert_freeze_F1}
  \begin{tabular}{cccccccc}
    \hline
    \bf{Project} & \bf{$f_0$} & \bf{$f_1$} & \bf{$f_2$}& \bf{$f_3$}&\bf{$f_4$}&\bf{$f_5$}&\bf{$f_6$}\\
    \hline
qt&0.36 & 0.35 & 0.35 & 0.35 & 0.35 & 0.35 & 0.35\\
openstack&0.44 & 0.41 & 0.41 & 0.41 & 0.41 & 0.41 & 0.41\\
jdt&0.64 & 0.65 & 0.64 & 0.63 & 0.64 & 0.64 & 0.64\\
platform&0.61 & 0.58 & 0.58 & 0.58 & 0.58 & 0.58 & 0.58\\
gerrit&0.21 & 0.2 & 0.2 & 0.2 & 0.19 & 0.2 & 0.2\\
go&0.61 & 0.56 & 0.56 & 0.56 & 0.56 & 0.56 & 0.56\\ \hline
mean & \textbf{0.48} & 0.46$\downarrow$ & 0.46$\downarrow$ & 0.46$\downarrow$ & 0.46$\downarrow$ & 0.46$\downarrow$ & 0.46$\downarrow$\\
  \hline
\end{tabular}
\end{table}
\textit{{\bfseries Finding 1}}: The first encoder layer of the BERT-style model plays the most important role in the downstream tasks.

\begin{figure}[h]
    \centering
    \subfigure[RoBERTaJIT]{
     \includegraphics[width=0.95\textwidth]{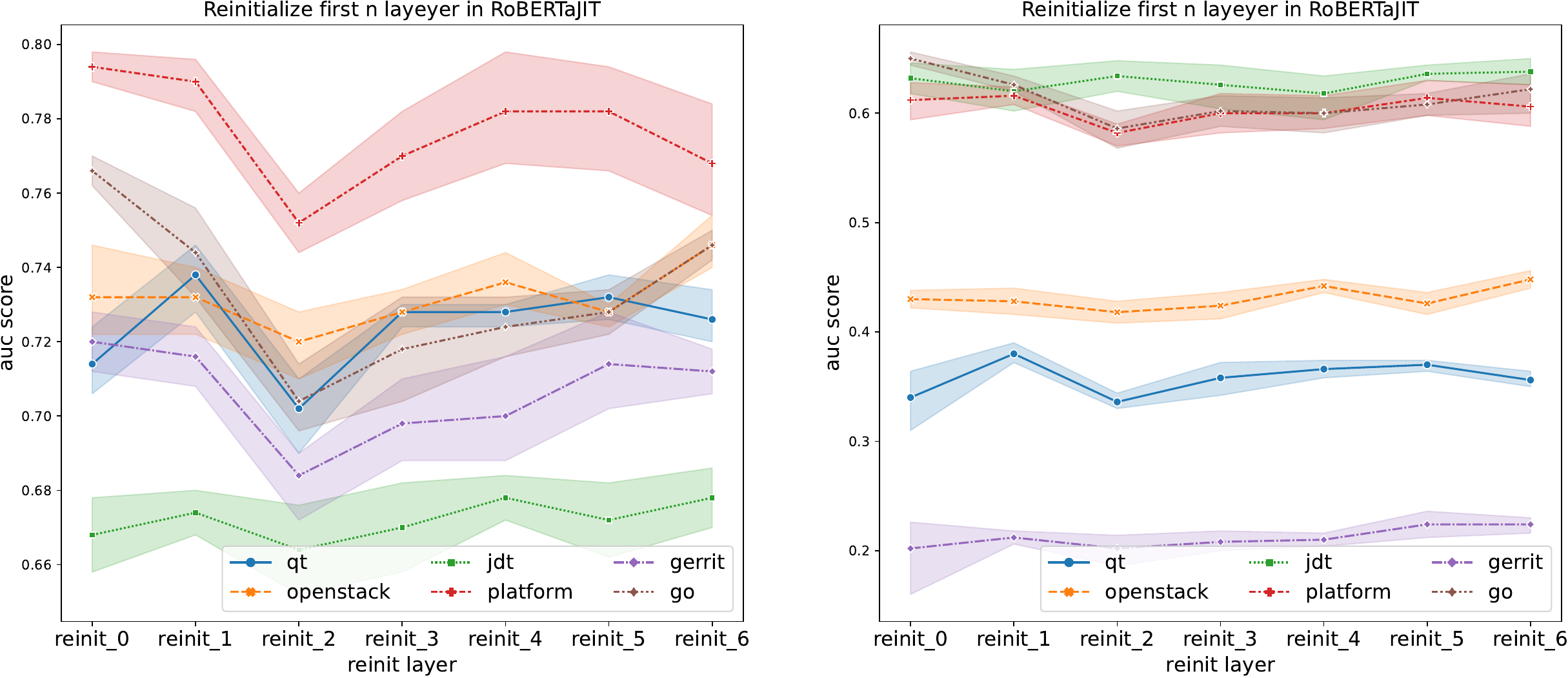}
     \label{fig:RoBERTa_reinit}
    }
    \subfigure[CodeBERTJIT]{
       \includegraphics[width=0.95\textwidth ]{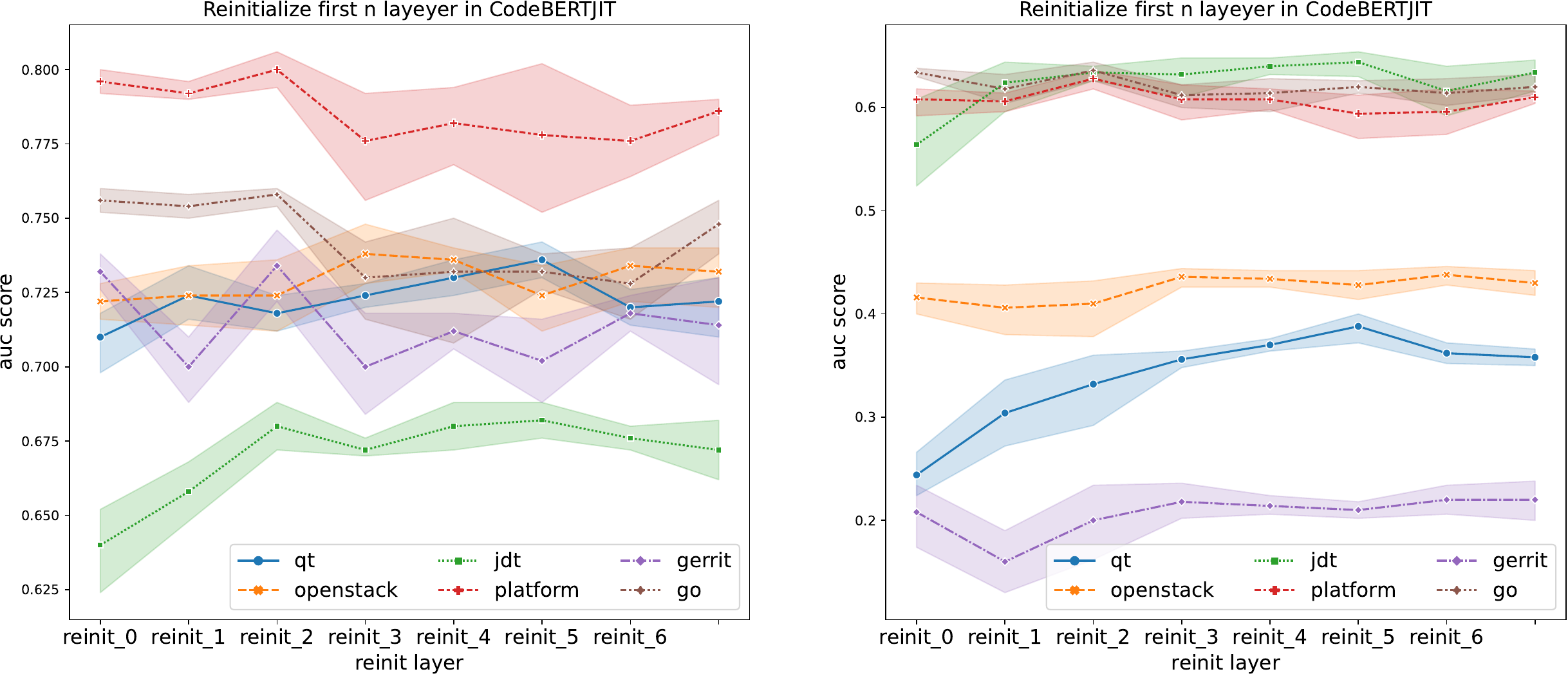}
       \label{fig:CodeBERT_reinit}
    }
    \caption{AUC and F1 score change on different parameter reinitialization settings for RoBERTaJIT and CodeBERTJIT}
    \label{fig:reinit}
\end{figure}

In another configuration of the model parameters, we initialize the first n layers of the pre-trained model, where n varies from 0 to 6. Employing the same initialization strategy as the original BERT model, we initialize the model weights using a normal distribution with parameters $\mathcal N (0,0.02^2)$. This strategy is associated with post-normalization design, where a smaller standard deviation proves more conducive to optimization. The experimental results for RoBERTaJIT and CodeBERTJIT are shown in Figure \ref{fig:reinit} considering the re-initialization of different encoder layers.

Although both RoBERTa and CodeBERT are BERT-style models, utilizing the same parameter initialization while keeping other settings frozen yields a distinct fluctuation trend in terms of model performance. In contrast to the performance decrease when freezing encoder layers in both models, parameter initialization demonstrates a slight improvement on some projects. In Figure \ref{fig:RoBERTa_reinit}, we observe an increase in AUC score for the \textit{qt} and \textit{jdt} projects when reinitializing the first encoder layer, with a corresponding improvement in the F1 score for the \textit{qt} project in RoBERTaJIT. Analyzing the results for CodeBERTJIT in Figure \ref{fig:CodeBERT_reinit}, it is evident that both AUC and F1 scores increase in the \textit{platform} and \textit{qt} projects when reinitializing the first two encoder layers. Furthermore, there is a performance increase in the \textit{qt} project when re-initializing the first four encoder layers. Initialization of model parameters does not necessarily lead to degradation of model performance, which is consistent with the conclusion of Zhang et al.'s work \cite{zhang2021revisitingbert}. The optimal value of n for parameter initialization varies across projects, offering insights into the necessity of adjusting different parameter initialization strategies during the fine-tuning of a downstream task, given the sensitivity of model performance. 

\textbf{\textit{Finding 2}}: A suitable initialization strategy customized for a project can improve the BERT-style model performance.

{\bfseries Results and analysis on RQ2.} In this research question, we aim at investigating the impact of different feature extractors in BERT-style JIT defect prediction models. Existing models adopt different feature extractors. For example, DeepJIT utilizes CNN as a feature extractor, CC2Vec employs LSTM, and LApredict employs logistic regression (implemented as a fully connected network). Additionally, transformer have emerged as a popular method for information extraction. Therefore, we select four distinct feature extractors: FCN (fully connected network), LSTM, transformer, and CNN. The experimental results for RoBERTaJIT and CodeBERTJIT using different feature extractors are provided in Table \ref{tab:roberta_extractors} and Table \ref{tab:codebert_extractors}.
\begin{table}\small
  \caption{Results of different feature extractors in RoBERTaJIT}
  \label{tab:roberta_extractors}
  \centering
  \begin{tabular}{ccccc}
\hline
    {\bf{Project}} & \bf{FCN} & \bf{LSTM} & \bf{transformer}& \bf{CNN}\\
\hline
    \bf{AUC score}& & & & \\
   qt & 0.74 &0.69&0.70&0.71\\
   openstack &0.75 &0.72&0.71&0.73\\
   jdt &0.61 &0.61&0.65&0.65\\ 
platform &0.79&0.77 &0.77&0.79\\
gerrit &0.72 &0.70&0.70&0.73\\ 
go &0.75&0.76&0.74&0.77\\ \hline
mean&\bf{0.73} &0.71&0.71&\bf{0.73}\\ \hline
\bf{F1 score}& & & & \\ 
   qt & 0.37 &0.32&0.23&0.30\\
   openstack &0.45 &0.42&0.42&0.44\\
   jdt &0.61 &0.57&0.61&0.61\\ 
platform &0.63&0.59 &0.61&0.62\\
gerrit &0.24 &0.21&0.18&0.25\\ 
go &0.60&0.63&0.63&0.59\\ \hline
mean&\bf{0.38} & 0.35 & 0.34 & 0.37\\ \hline
\end{tabular}
\end{table}

We conducted a comparison of AUC scores and F1 scores for both RoBERTaJIT and CodeBERTJIT across six projects with different feature extractors. The mean AUC scores ranged from 0.71 to 0.73 in RoBERTaJIT and 0.71 to 0.75 in CodeBERTJIT, while the mean F1 scores varied from 0.34 to 0.38 in RoBERTaJIT and 0.45 to 0.49 in CodeBERTJIT. Notably, the poorest performance occurred when utilizing the transformer, while the best performance was observed with FCN. 

\textit{{\bfseries Finding 3}}:  Complex feature extractor will lead to poor model effectiveness. For the BERT-style defect prediction model, using FCN for feature extraction can achieve great performance.
\begin{table}[h]\small
  \caption{Results of different feature extractors in CodeBERTJIT}
  \centering
  \label{tab:codebert_extractors}
  \begin{tabular}{ccccc}
    \hline
    {\bf{Project}} & \bf{FCN} & \bf{LSTM} & \bf{transformer}& \bf{CNN}\\
    \hline
     \bf{AUC score}& & & & \\
   qt & 0.73 &0.73&0.71&0.74\\
   openstack &0.75 &0.75&0.73&0.73\\
   jdt &0.68 &0.66&0.65&0.68\\ 
platform &0.82&0.79 &0.78&0.80\\
gerrit &0.75 &0.72&0.67&0.70\\ 
go &0.76&0.75&0.72&0.76\\ \hline
mean&\bf{0.75} & 0.73 & 0.71 & 0.74\\ \hline
\bf{F1 score}& & & & \\
 qt & 0.37 &0.37&0.34&0.37\\
   openstack &0.45 &0.43&0.43&0.43\\
   jdt &0.65 &0.63&0.55&0.65\\ 
platform &0.61&0.62 &0.60&0.62\\
gerrit &0.20 &0.23&0.17&0.20\\ 
go &0.64&0.64&0.59&0.64\\ \hline
mean&\bf{0.49} & \textbf{0.49} & 0.45 & \textbf{0.49}\\ 
  \hline
\end{tabular}
\end{table}
\begin{table}[h]\small
  \caption{Results of RoBERTaJIT with different weight decay rate}
  \centering
  \label{tab:roberta_weight_decay}
  \begin{tabular}{cccccccc}
\hline
  \bf{weight decay rate}& {\bf{qt}} & \bf{openstack} & \bf{jdt} & \bf{platform}& \bf{gerrit}&\bf{go}&\bf{mean}\\
    \hline
     \bf{AUC score}& & & & \\
0&0.72 & 0.74 & 0.67 & 0.8 & 0.7 & 0.76 & 0.73 \\
 1e-5&0.72 & 0.74 & 0.66 & 0.79 & 0.74 & 0.75 & 0.73  \\
1e-4 &0.70 & 0.74 & 0.65 & 0.81 & 0.74 & 0.75 & 0.73  \\
1e-3 &0.71 & 0.74 & 0.68 & 0.81 & 0.76 & 0.78 & \textbf{0.75} $\uparrow$  \\ \hline
\bf{F1 score}& & & & \\
0&0.27 & 0.36 & 0.66 & 0.56 & 0.17 & 0.57 & 0.43  \\
1e-5&0.28 & 0.35 & 0.66 & 0.56 & 0.17 & 0.58 & 0.43  \\
1e-4&0.27 & 0.35 & 0.66 & 0.56 & 0.17 & 0.57 & 0.43  \\
1e-3&0.29 & 0.37 & 0.6 & 0.56 & 0.17 & 0.58 & 0.43  \\
  \hline
\end{tabular}
\end{table}

{\bfseries Results and analysis on RQ3.} In this research question, we aim to explore the effectiveness of different hyperparameter settings in the Adam optimizer. The selection of an optimizer is crucial, which can decide the update strategy of model weights. In the original code implementation of RoBERTaJIT and CodeBERTJIT, the optimizer selected is Adam, and the optimizer used in RoBERTa and CodeBERT is AdamW, which utilizes weight decay in Adam. Adam is an adaptive learning rate optimization algorithm using momentum and scaling. The optimizer is designed to be suitable for non-stationary targets and problems with very noisy and/or sparse gradients. The parameter update strategy of Adam can be expressed as the following equation. 
\begin{equation}
    \boldsymbol{\theta}_{t+1}=(1-\lambda) \boldsymbol{\theta}_{t}-\alpha \nabla f_{t}\left(\boldsymbol{\theta}_{t}\right)
\end{equation}

During the training process, the weight vector can be represented by $\theta$, and the learning rate can be represented by $\alpha$, and $\nabla f_{t}\left(\boldsymbol{\theta}_{t}\right)$ is the gradient of loss function at time step $t$. $\lambda$ represents a certain value of the weight decay rate. 
$\lambda$ in Adam is equal to zero and other non-zero values in AdamW. Here we set $\lambda \in\{0,1e-5,1e-4,1e-3\}$, which is the commonly selected value when training a model. We conducted experiments on six projects. Table \ref{tab:roberta_weight_decay} and Table \ref{tab:codebert_weight_decay} show the results of the two models with different weight decay rates.
\begin{table}\small
  \caption{Results of CodeBERTJIT with different weight decay rate}
  \centering
  \label{tab:codebert_weight_decay}
  \begin{tabular}{cccccccc}
    \hline
  \bf{weight decay rate}& {\bf{qt}} & \bf{openstack} & \bf{jdt} & \bf{platform}& \bf{gerrit}&\bf{go}&\bf{mean}\\
    \hline
     \bf{AUC score}& & & & \\
0&0.72 & 0.71 & 0.66 & 0.79 & 0.7 & 0.77 & 0.73  \\
1e-5&0.74 & 0.72 & 0.67 & 0.78 & 0.7 & 0.76 & 0.73  \\
1e-4&0.74 & 0.72 & 0.66 & 0.78 & 0.71 & 0.76 & 0.73  \\
1e-3&0.72 & 0.75 & 0.65 & 0.79 & 0.69 & 0.75 & 0.73  \\ \hline
\bf{F1 score}& & & & \\
0&0.26 & 0.39 & 0.61 & 0.61 & 0.18 & 0.62 & 0.45  \\
1e-5&0.35 & 0.43 & 0.65 & 0.61 & 0.2 & 0.64 & \textbf{0.48} $\uparrow$  \\
1e-4&0.38 & 0.43 & 0.62 & 0.61 & 0.22 & 0.59 & \textbf{0.48} $\uparrow$ \\
1e-3&0.35 & 0.39 & 0.54 & 0.62 & 0.19 & 0.64 & 0.46 $\uparrow$  \\
  \hline
\end{tabular}
\end{table}

When we add weight decay strategy, the mean performance of six models is non-decreased or slightly improved when $\lambda =1e-3$ in RoBERTaJIT and $\lambda \in \{1e-5,1e-4,1e-3\}$ in CodeBERTJIT. Some other researchers have explored that weight decay can make fine-tune more stable in few-shot fine-tuning of RoBERTa, our experiments extend to BERT-style JIT defect prediction models and the results reveal the weight decay strategy is still effective in BERT-style model under the full fine-tuning scenario. 

\textit{{\bfseries Finding 4}}: AdamW can be used to slightly improve the performance of BERT-like model.

{\bfseries Results and analysis on RQ4.} In this research question, we aim to explore if there exists a method to fine-tune a pre-trained model and keep the performance non-decreasing more efficiently. In RQ1 to RQ4, we have explored some factors which can affect model performance in JIT defect prediction task. And LoRA can help us fine-tune a pre-trained model with lower resource utilization. We customized LoRA with our findings to find a cost-effective fine-tune solution. We do not freeze the first encoder layer based on the findings of RQ1 and using the weight decay strategy to update LoRA weights, we select FCN as the feature extractor, Table \ref{tab:optimized_method} shows the results of the customized fine-tune method on top of LoRA. 

\begin{table}\small
  \caption{AUC results of OCJITLoRA compare to baselines}
  \label{tab:optimized_method}
  \centering
  \begin{tabular}{ccccccc}
  \hline
    {\bf{Project}} &\textbf{LApredict}& \bf{CodeBERTJIT} & \bf{OCJITLoRA} &\textbf{DeepJIT} &\textbf{CC2Vec } \\
    \hline
   qt& 0.74& 0.73 &0.76&0.70& 0.69\\
   openstack &0.74&0.74 &0.74&0.72& 0.73\\
   jdt &0.68&0.67 &0.68&0.67& 0.68\\ 
platform&0.75 &0.80&0.81&0.77& 0.77 \\
gerrit &0.75&0.72 &0.72&0.70& 0.70 \\ 
go &0.68&0.76 &0.76& 0.69& 0.69 \\ \hline
mean&0.72&0.74 &\bf{0.75}& 0.71 & 0.71\\ 
  \hline
\end{tabular}
\end{table}

\begin{table}\small
  \caption{F1 results of OCJITLoRA compare to baselines}
  \label{tab:optimized_method}
  \centering
  \begin{tabular}{ccccccc}
  \hline
    {\bf{Project}} &\textbf{LApredict}& \bf{CodeBERTJIT} & \bf{OCJITLoRA} &\textbf{DeepJIT}& \bf{CC2Vec}\\
    \hline
   qt& 0&0.36&0.38&0.33& 0.29\\
   openstack& 0.13&0.44&0.43&0.41& 0.43\\
   jdt &0.42&0.61&0.64&0.64& 0.65\\ 
platform&0.09&0.61&0.60&0.58& 0.56\\
gerrit &0.09&0.21&0.22&0.19& 0.17\\ 
go &0.06&0.61&0.63&0.57& 0.58\\ \hline
mean&0.13&0.47&\bf{0.48}&0.45&0.43\\ 
  \hline
\end{tabular}
\end{table}
As shown in Table \ref{tab:optimized_method}, we compare the results of the fine-tuned CodeBERT model based on customized LoRA (called OCJITLoRA) with vanilla CodeBERTJIT and other three efficient mehtods DeepJIT, CC2Vec and LApredict. We can see that OCJITLoRA can achieve best performance among these models according mean results. Compare to CodeBERTJIT, although improvement of OCJITLoRA is small, it is worth mentioning that CodeBERTJIT need nearly 21 Gib memory to train a model but OCJITLoRA only need nearly 7Gib memory. Which means memory consumption is reduced to one-third of the CodeBERTJIT and achieve comparable performance. 

\textit{{\bfseries Finding 5}}: By combining some optimization factors and customizing LoRA, the performance of the CodeBERT can be improved with lower resource consumption.
\section{Threats to Validity}
\label{sec:threats}
The main limitations in our work mainly come from the information loss caused by truncation. 

Limited by the max input length which is equal to 512 of BERT-style models, our work conducts truncation operation on the commit code and commit message longer than 512. Such operation brings information loss situation which also appears in other work. Although there are many researchers have explored extending input length in pre-trained models, memory consumption and computing time are still a holdback from dealing with long code or text sequences. How to ensure the integrity of the original input information is a significant problem.
\section{Conclusions}
\label{sec:conclusions}
We have conducted a detailed empirical analysis of the different fine-tuning strategies that affect model performance. First, We demonstrate that BERT-style-based models have a robust performance in extending big-scale datasets. Second, we conduct parameter freezing experiments and find the first encoder layer of pre-trained models plays a significant role can deciding the upper bound of model performance. Different from the decreased performance brought by parameter freezing, proper weight parameters initialization of the model can improve the performance of JIT defect prediction. Third, we demonstrate that for parallel structure data like different commit code patches, a simple feature extraction operation like FCN and CNN can bring good performance in BERT-style-based models. Fourth, we explore the effect of the weight decay strategy used in the Adam optimizer and demonstrate that the weight decay strategy can bring a slight improvement in model performance. Finally, we combine our findings and a parameter-efficient tuning method LoRA to optimize the process of tuning in CodeBERTJIT and obtain a slight improvement in six projects under one-third memory consumption than original tuning process. 
\bibliographystyle{splncs04}

\end{document}